# VIRTUAL BACKBONE TREES FOR MOST MINIMAL ENERGY CONSUMPTION AND INCREASING NETWORK LIFETIME IN WSNS

Saibharath S[1] and Aarthi.J[2]

[1]Department of Computer Science BITS Pilani, Hyderabad Campus Hyderabad.India
[2]Department of Computer Technology, MIT Campus, Anna University, Chennai, India

## ABSTRACT

*Virtual backbone trees have been used for efficient communication between sink node and any other node in the deployed area. But all the proposed virtual backbone trees are not fully energy efficient and EVBTs have few flaws associated with them. In this paper two such virtual backbones are proposed. The motive behind the first algorithm, Most Minimal Energy Virtual Backbone Tree (MMEVBT), is to minimise the energy consumption when packets are transmitted between sink and a target sensor node. The energy consumption is most minimal and optimal and it is shown why it always has minimal energy consumption during any transfer of packet between every node with the sink node. For every node, route path with most minimal energy consumption is identified and a new tree node is elected only when a better minimal energy consumption route is identified for a node to communicate with the sink and vice versa. By moving sink periodically it is ensured the battery of the nodes near sink is not completely drained out. Another backbone construction algorithm is proposed which maximises the network lifetime by increasing the lifetime of all tree nodes. Simulations are done in NS2 to practically test the algorithms and the results are discussed in detail.*

## 1. INTRODUCTION

Wireless sensor networks (WSN) can be defined as the number of sensor nodes which form a network and monitor the location and send the data to the sink. Some of the main applications of WSN are machine health monitoring, smart home monitoring, area monitoring etc. To achieve low cost monitoring, sensors are deployed in the specific area to be monitored. These sensors communicate with each other through wireless link. All the nodes must send their messages to the sink node in an energy efficient manner. For sustenance of the network, one must have both energy efficient transmission and high overall network lifetime. The main goal is to minimize the energy consumed and maximize the network lifetime. Clustering and backbone formation are two contrasting schemes for efficient routing in wireless sensor network [13-14]. Various types of clustering like distributed clustering, intra cluster routing exists. Basically every node will transfer its message to its cluster head and the cluster head would communicate to sink via many intermediate cluster head. Virtual backbone tree is a backbone formed to cover the entire network. Backbone consists of tree nodes which are selected based on the energy of the sensor node, fitness factor, distance from the sink node and also the angle in which the node has to transmit and receive packet from its parent. Other sensor nodes receive and send messages via the tree nodes. Every sensor node(non-tree node) will select a tree node as its parent. Virtual backbones not only provide energy efficient communication infrastructure but increases the network lifetime. The tree nodes which are near the sink node are prone to failure soon as most of the messages which are communicated pass through them. The biggest issue with the sensors is their limited power. In this paper, sink is assumed to be a mobilizer which has the ability to move within the sensor network and it has unlimited power.

## 2. RELATED WORKS

Virtual backbone tree has been constructed in many ways. One of the approaches used is by constructing connected dominating set [8-10]. T.Acharya et al build a connected dominating set which is distributed and nodes are given status as tree nodes dynamically and they forward their data to their parent and node does not need any routing table [1]. H. Raei et al have constructed a virtual backbone by first building a maximal independent set and then connecting them together [2]. Connected dominating set (CDS) gives minimal number of nodes through which a backbone is formed. When large number of packets is transferred through these nodes, energy of these backbone nodes get drained quickly leading to more re-construction and sometime network failure. Hence multiple connected dominating set were formed. Jing He et al have proposed a load balanced CDS, in which workload is shared between nodes and the number of dependents of particular tree node is load balanced [3]. Zhao et al maximize the overall network lifetime by constructing virtual backbone through rules 1, 2 and k and have implemented virtual backbone through iterative local replacement [4].

The second approach is to send BCR packet from the sink. The node which receives it calculates its fitness factor with respect to the node which had sent it. Time delay is calculated based on the fitness factor. Before this delay expires, if the node receives another BCR request packet it becomes a sensor node i.e. a non tree node else it becomes a tree node and becomes a part of the backbone. The non tree node picks the nearest tree node as its parent [5]. ViTAMin [6] is an extension of EVBT proposed in [5]. For all non tree node instead of picking the nearest tree as its parent, for each tree node distance from the sink is stored and a modified EBCR request packet is sent through with every non tree node is associated with a parent tree node such that it has a shortest path to the sink [6]. In modified-EVBT, energy consumed through various links are tracked in the every node and forwarded to their child node leading to a better EVBT in terms of energy consumed.

## 3. PROPOSED WORK

### 3.1 Most Minimal Energy consuming Virtual Backbone Tree (MMEVBT)

In this MMEVBT, firstly backbone tree construction algorithm has been proposed which ensures minimal energy consumption for all the packets sent by all the nodes. Though the energy consumption is kept minimal, the energy of tree nodes nearer to the sink will decrease rapidly. To overcome this, the sink which is a mobilizer changes its position periodically. In few applications like monitoring military battle fields, sink cannot move to any position in what wants to. So sink places itself in the best possible position which it can.

Based on the energy of a node, the node is given a status. If the energy is greater than threshold it is eligible to become a tree node but there are other conditions as well. A node can become a tree node when some node wants it as parent based on energy consumption between sink and itself. The main objective is to select a route path for every node which must be the most minimal and optimal path in terms of energy consumption to send a packet to the sink.
Always a $node_i$ is a tree node when a $node_j$ wants it to be its parent. Any $node_j$ which is a child of $node_i$ can elect some other node as its parent $node_k$ when energy consumption through $node_i$ to the sink node is becoming greater than $node_k$. If a tree $node_i$ which had child/children but now had lost all of them by the above process, then the $node_i$ becomes a non tree node which can become a tree node later on.

Each node has one of the statuses associated with it: Tree node, Permanent non tree node, Non tree node which can become tree node or a failed node. Total energy needed to send a packet from a node is minimal and optimal.

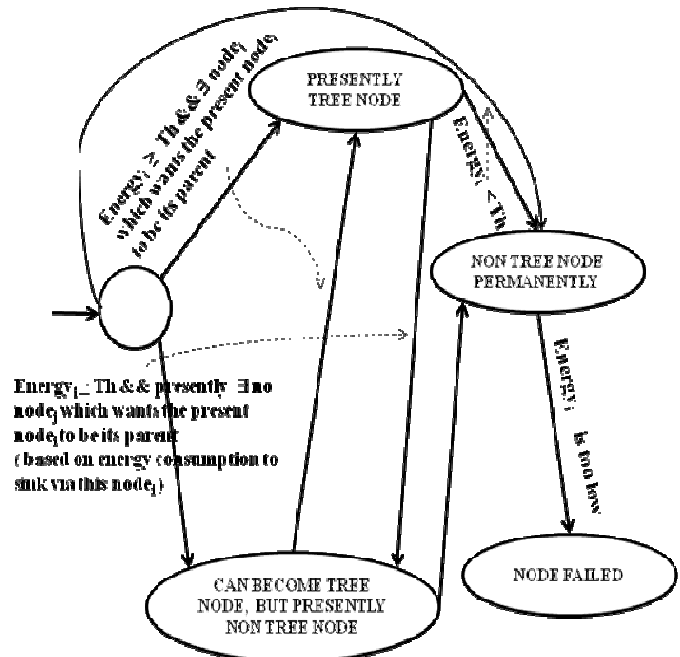

Figure 1. State diagram showing the change in status of a node

if $Energy_i \geq Th$ && Count(No. child($node_i$))<=0
   STATUS of node: Non tree node which can become tree node if some node wants it
              to be parent
else if $Energy_i \geq Th$ && Count(No. child($node_i$))>0
   STATUS of node: Tree node
else if $Energy_i < Th$
   STATUS of node: Non Tree node
else
   STATUS of node: Node failed

The energy consumption in data delivery is minimal because:

i)       All nodes whose energy greater than threshold and has the ability to be a tree node are selected
ii)      For every node N,
a.       One finds the shortest path to the sink based on energy consumption
b.       For all nodes which can become parent, whichever has Energy consumption (Sink→ Parent + Parent → Node N) as minimum, then this parent becomes the tree node if not before and a parent node for N.
c.       Always this above shortest path is updated and a tree node can become a non tree node (with can become tree node as its status)
d.       Whenever a link failure happens, the above algorithm is continuously updated.

A route with the minimum energy consumption is always selected to send a packet between sink and a particular node.

Periodically sink changes its position so that the tree nodes nearby it don't drain out. Sink finds the region of nodes having more average battery power and moves there. Moving sink periodically ensures the overall network lifetime is increased.

### 3.2 Increasing network lifetime by reducing node failures and reconstruction in WSNs

Another backbone construction algorithm has been proposed which increases overall network lifetime. In this a child node can select one from multiple tree nodes to forward data to sink. In existing VB construction mechanism best tree node was picked to be its parent. But when the node sends many messages frequently, the tree node's energy can be drained quickly. Though there are many other tree nodes as it chooses its parent which is best tree node, the energy of its parent is drained. To avoid this situation, this construction mechanism has been proposed.

Nodes which can be tree nodes are initially picked if it has energy greater than the threshold. Every node has a fitness factor with respect to another node based on the distance, energy of the node and the angle between the nodes.

Fitness factor $f_i$(node n) is the fitness factor of node i with respect to the node n.

$$f_i(n) = c_1 f_d + c_2 f_e + c_3 f_\beta \qquad (1)$$

where $c_1+c_2+c_3=1$. $f_d$ is the inverse of distance between $node_i$ and n. $f_e$ is the energy of node n. $f_\beta = \frac{\pi}{|\beta|}$, where β is the angle between the two nodes. It must be kept straight to ensure better fitness factor.

Every node can send the data to sink through any one of the tree nodes. Let $t_{i0}, t_{i1}, t_{i2} \ldots t_{i(k-1)}$ to indicate which one of the immediate tree nodes(totally k nodes from 0…k-1 ) to be selected to send the data to sink. Thus $t_{i0}$ would be 1 if the node $node_0$ is selected to send the data. Likewise $t_{ip}$ would be 1 if the $node_p$ is selected to send the data where p<k. The maximum number of child nodes which can choose a tree node to send its data must be minimized to ensure increase in overall network lifetime. If a particular node with higher fitness factor is used to send the data to sink by many nodes, then it is prone to failure as the energy would be drained alarmingly. Let count(i) denote the number of child nodes select the tree $node_i$ to send its data. Let $mc = max_i \; count(i)$ be the maximum number of nodes which select any tree $node_i$. Our aim is to minimize mc.

For each tree node T which can be selected by any child node to send their data,

$S_T = \{i | node_i \; sends \; to \; tree \; node \; T \; if \; t_{iT} = 1\}$ One can state the problem as,

$$Minimize \; mc \qquad (2)$$

$$\forall i \; t_{i0} + t_{i1} + \cdots t_{i(k-1)} = 1 \qquad (3)$$

$$\forall t \; \sum_{i \in S_T} t_{iT} \leq mc \qquad (4)$$

$$\forall i \; t_{i0}, t_{i1}, \ldots t_{i(k-1)} \in \{0,1\} \qquad (5)$$

The objective is to minimize mc (2).(3) ensures the packet is sent to only one of the tree node and can be different for each node$_i$.(5) ensures at most mc nodes send through tree node T. One of the $t_{i0}, t_{i1}, \ldots t_{i(k-1)}$ is 1, but choosing this becomes difficult. A node with respect to another node can be judged based on the fitness factor.

In figure 2, with respect to node $N_7$, $f_i(N_4)=0.58$ and $f_i(N_5)=0.62$. Now for the node $N_7$ to communicate with the sink it must select either one of Node $N_4$ or $N_5$. Instead of permanently selecting one of the tree node it is done in the following way.

Probability of sending the packet via node $N_4$ = $\frac{f_i(N_4)}{\sum_k fitness\ factor\ f_i(N_K)}$ where k refers only to the nodes which are competing to become the parent of node $N_7$. Here in this algorithm there is no fixed parent, the packet is sent to the sink only based on the probability. In other words here, Probability of sending via node $N_4$ = $\frac{fi(N_4)}{fi(N_4)+fi(N_5)}$. Similarly the probability of sending the packet via node $N_5$ = $\frac{f_i(N_5)}{f_i(N_4)+f_i(N_5)}$.

Suppose N5 is selected, it is having access via two tree nodes N2 and N3. Now again it takes the route path in any of the possibilities with probability of taking be $\frac{f_i(N_2)}{f_i(N_2)+f_i(N_3)}$ and $\frac{f_i(N_3)}{f_i(N_2)+f_i(N_3)}$ respectively. Independently for each node$_i$ which needs to send one of a tree node T out of k tree nodes,

$$P(t_{iT} = 1) = \frac{fi(N_T)}{\sum_k fitness\ factor\ fi(N_K)} \tag{6}$$

Now to select a tree node T to send one can select random number R between 0 and 1. If $R \in [0, \frac{f_i(N_0)}{\sum_k fitness\ factor\ f_i(N_K)}]$, then data is sent to tree node$_0$ out of k tree nodes. If $R \in (\frac{f_i(N_0)}{\sum_k fit.factor\ f_i(N_K)}, \frac{f_i(N_1)}{\sum_k fit.factor\ f_i(N_K)}]$, then data is sent to tree node$_1$ out of k tree nodes. Similarly tree node is chosen to send to the message.

$$E[count(i)] = \sum_{i \in S_T} E[t_{iT}] \tag{7}$$

$$= \sum_{i \in S_T} (1.P(t_{iT} = 1) + 0.P(t_{iT} = 0)) \tag{8}$$

$$= \sum_{i \in S_T} (P(t_{iT} = 1)) \tag{9}$$

$$E[count(i)] = \sum_{i \in S_T} \frac{f_i(N_T)}{\sum_k fit.factor\ f_i(N_K)} \leq mc \tag{10}$$

Expectation of number of nodes which chooses tree node$_i$ is calculated in (7),(8),(9) and (10). One can apply in markov and chebbyshev's inequalities which guarantee that in any distribution, nearly all values are close to the mean.

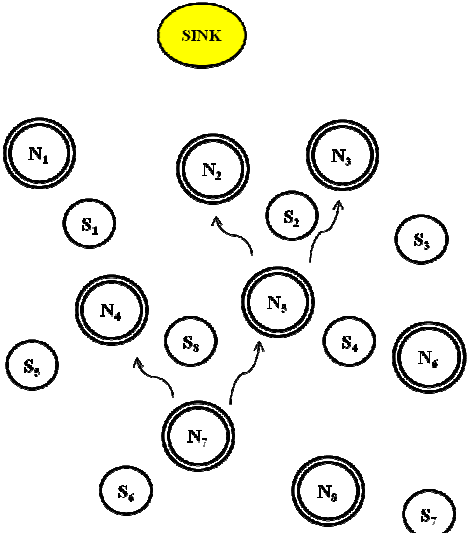

Figure 2. Node $N_7$ can send a packet through N4 or N5

All upstream links through the tree nodes based on the fitness factor with respect to the other node to which it communicates. Probability of choosing to send via a particular tree node is directly proportional to its fitness factor. Though one can choose any path, there is a higher chance of taking the path via the node which has highest fitness factor. Here in this algorithm one tries to use as many links as possible. Suppose one always concentrates on one path, the energy level of particular node which is elected as a tree node get depleted. By using only one particular tree node to communicate always one can reduce energy consumption. But sooner than later its energy gets drained out by sending too many packets, link can get busy which can lead to overloading the particular node. When there is a failure of this node the tree gets broken and it is tough to reconstruct as more nodes use this as their parent. This node has every possibility to become a permanent non tree node soon. By using this scheme overloading of particular tree node is avoided and depending upon the capacity(fitness factor) of a tree node it is allotted packets to be sent.

## 4. SIMULATION

### 4.1 Construction of backbone in MMEVBT

Here a tree node initiated scheme from the sink for the minimal energy consumption is discussed. Initially all nodes with energy ≥ Th act as a non tree node but it can become tree node if a child node elects it. If node's energy < Th, it becomes non tree permanently.

First nodes which can sense sink in their range finds the energy consumption from the node to sink (direct transfer). The other nodes find the energy consumption from it to the node n ( if node n is in its sensing range) which already has direct access to sink. Say a node k has several nodes which has direct access to sink in its sensing range. It selects the node with which it has minimum E(k,n) + Consumption(n)  value. The node k elects the node n to be a tree node. Note node n must have energy greater than the threshold. Consumption of node k is defined as,

$$Consumption(k) = E(n,k) + Consumption(n)$$

The same process is repeated for all nodes. Nodes elect a node as a tree node if it already has a path to the sink and also it must have the minimum E(k,n) + Consumption(n). Here minimal energy consumption is ensured for any packet sent from a particular node to the sink and vice versa by making child nodes electing the necessary tree nodes.

Simulation is done in NS2. In figure 3, for each sensing range all the three algorithms are made to run with the same random deployment of nodes. The average number of tree nodes is calculated through 15 successful formation of single virtual backbone tree in each sensing range. For d=20 m, no single virtual backbone tree could be formed after several attempts. For d=25m, 15 successful single VBT was captured after 23 failures. For d=30m and 35m, 3 and 1 failures occurred respectively.

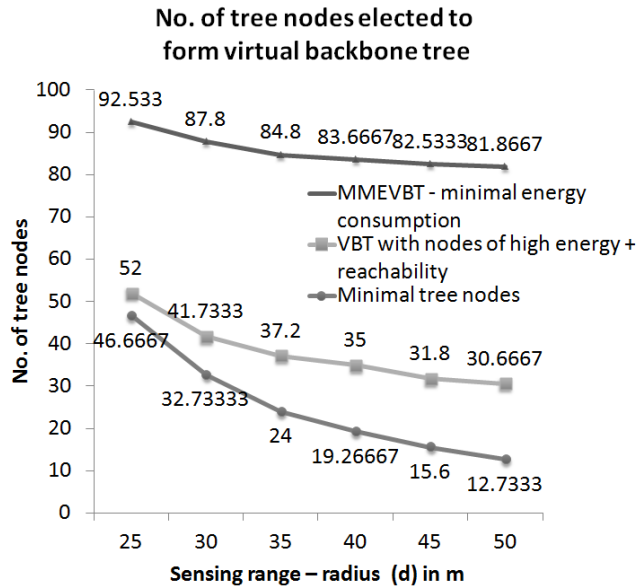

Figure 3. No. of tree nodes when 200 nodes deployed in square field of 200m x 200m

### 4.2 VBT with Minimal Tree Nodes

Initially node with highest reachability is picked. Nodes within its sensing range are covered. Covered$_i$ with value 1 and 2 indicates it covered and it is a tree node respectively. Next in the nodes which are covered, one finds the node which has maximum number of reachable nodes in |Total nodes – Covered nodes|. Such a node is made a tree node. This process goes on till all nodes are covered.

```
Set max -1
For each node i
        covered_i =0
        if max < count(rts_i)
                max= count(rts_i); maxid=tr;
covered_tr=2; flag=true;
//when covered_i=2 indicates it is picked as tree node
for each node k
        If Nodek.rn(tr)==true
                covered_k=1;
while(flag)
```

```
   for each node b
        wd_b=0
 for each node b
        if covered_b==1 && energy ≥ Th
           for each node c
                if covered==0&&Node_b.rn(c)==true
                        wd_b=wd_b+1
max=-1
for each node b
 if wd_b>max
        max=wd_b;maxid=b;
covered_maxid=2 // pickes as tree node
flag=false
For each node k
        If Node_k.rn(maxid)==true && covered_k!=2
                covered_k=1;
        else if covered_k==0
                flag=true
end
```

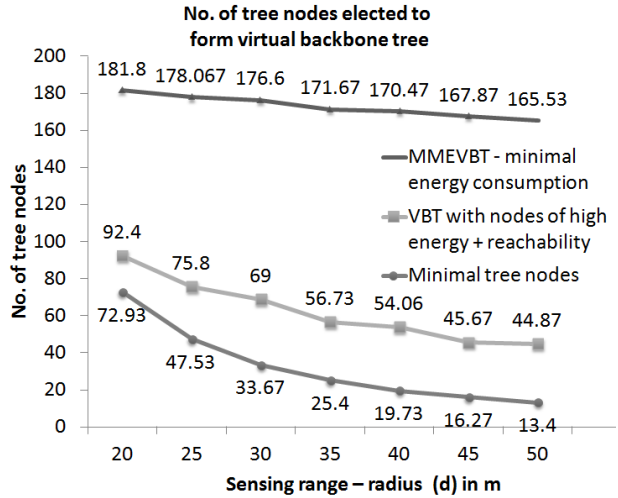

Figure 4. No. of tree nodes when 400 nodes deployed in square field of 200m x 200m

Figure 4 depicts the number of tree nodes elected to build a single virtual backbone tree. For d=15 m, no single VBT could be formed. For d=20m, 4 unsuccessful attempts out of 19 occurred.

| Th | Threshold |
|---|---|
| $Energy_i$ | Energy of node i |
| Consumption(i) | Total energy consumed in the nodes from node i to the sink. When a packet is sent. |
| Energy(k,n) | Energy consumed in node k when trying to send a packet to node n. |
| Fi(n) | Fitness factor of node I with respect to node n |

Table 1. Notations used in the paper

## 5. CONCLUSION

In this paper virtual backbone algorithm with minimum energy consumption has been discussed. This algorithm ensures always most minimal energy is always consumed when a message is sent from any node to the sink. Another algorithm which increases the network lifetime has been discussed. This is not deterministic algorithm. The expected number of nodes which chooses a particular node to forward the data is calculated and number of messages passing through a node is minimized.

**Authors**

Saibharath received B.E degree in computer science and engineering from MIT Campus,AnnaUniversity in 2013. He is currently pursuing M.E degree in Computer science in BITS Pilani, Hyderabad campus, India. His currentresearch area lie in WSN and cyber forensics in cloud computing.

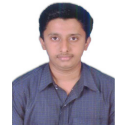

Aarthi is an undergraduate student in the Department of Computer Technology at MIT Campus, Anna University, India. Her current research interest lie in the area of wireless sensor networks.

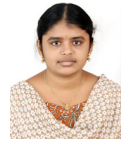